\documentstyle[preprint,aps]{revtex}
\newcommand{\be}{\begin{equation}}
\newcommand{\ee}{\end{equation}}
\newcommand{\bd}{\begin{displaymath}}
\newcommand{\ed}{\end{displaymath}}
\newcommand{\baa}{\begin{array}{lll}}
\newcommand{\eaa}{\end{array}}
\newcommand{\ba}{\begin{eqnarray}}
\newcommand{\ea}{\end{eqnarray}}

\begin{document}
\draft

\title{ Gluon polarization in transversely polarized nucleons
and jet spin asymmetries  at RHIC}
\author{J. Soffer \cite{1}}
\address{Centre de Physique Th\'eorique - CNRS - Luminy,\\
Case 907 F-13288 Marseille Cedex 9 - France}
\author{and\\ O. V. Teryaev\cite{2}}
\address{Bogoliubov Laboratory of Theoretical Physics, \\
Joint Institute for Nuclear Research, Dubna, 141980, Russia}
\maketitle

\begin{abstract}
We derive the analog of the Wandzura-Wilczek relation for the light-cone
distributions of polarized gluons in a transversely polarized nucleon.
The short distance cross-section is entirely due to the intrinsic
transverse momenta of gluon in the nucleon, in complete analogy
to the quark case. Numerical estimate for the double spin transverse
asymmetries at RHIC are presented.
\end{abstract}
\pacs{PACS. 12.38.Bx, 13.88.+e}

\narrowtext

The spin properties of gluons and quarks are rather different. In particular,
this is manifested in the fact, that there is no analog of the twist-two
transversity distributions for gluons and  their contribution
to transverse asymmetry starts at twist-three level. This is leading,
generally speaking, to the relative suppression
of gluon transverse asymmetries with respect to the quark ones,
which was used recently to formulate the selection rule \cite{JaSa}
in QCD, which is of direct relevance for the physics programme
of the future polarized $pp$ collider at RHIC.

However, the detailed analysis of the quark contribution to the
double transverse spin asymmetries $A_{TT}$
using the Monte-Carlo simulation \cite{MS}, resulted in rather small numbers
of the order of $1\%$, and therefore it seems natural to
question the role of gluon corrections.
This is a subject of the present paper.

The transverse polarization effects are arising from two basic
sources: the leading twist transversity distribution, resulting
in the correlation of transverse polarizations, and the twist-three
parton correlations, suppressed by hadron mass. While the first are
absent for gluons, the gluon correlations are, generally speaking,
rather complicated \cite{Ji}.
At the same time, the experimental data on the $g_2$ structure
function \cite{g2}
do not deviate strongly from the twist-two approximation,
suggested by Wandzura and Wilczek (WW) \cite{WW}, whose physical meaning
is just the dominance of the effect of transverse motion of the quark
over that of the gluon field \cite{kt}.
This is the reason, why we are suggesting here
a generalization of the WW approximation  to
the case of gluons.

To do this, let us start from the light-cone density matrix of gluon,
where we ignore twist-four term, namely:

\begin{eqnarray}
\label{dm}
\int{{d\lambda \over2\pi}}
e^{i\lambda x}
\langle p,s|A^\rho(0) A^\sigma (\lambda n) |p,s \rangle =
d^{\rho \sigma}G(x)+M[\Delta G (x) (sn) \epsilon ^{\rho \sigma p n}
+\Delta G_T (x)
\epsilon ^{\rho \sigma s_T n}],
\end{eqnarray}
where $n$ is the gauge-fixing light-cone vector such that $np=1$,
and we define the two transverse tensors
$d_{\rho \sigma}=g_{\rho \sigma}-p_\rho n_\sigma-n_\rho p_\sigma$ and
$\epsilon ^{\rho \sigma p n}=\epsilon ^{\rho \sigma \alpha \beta }
p_{\alpha }n_{\beta }$. We denote by
$s_{\mu }$ the covariant
polarization vector  of the proton of momentum
$p$ and mass $M$ and we have $s^2=-1, sp=0$ and $s_T=s-p(sn)$ which
corresponds to
the transverse polarization.
Here $G(x)$ and $\Delta G(x)$ are the familiar unpolarized gluon distribution
and gluon helicity distribution, respectively.
The transverse gluon distribution $\Delta G_T$
is the most natural measure of transverse polarization,
analogous to the quark structure function $g_T=g_1+g_2$,
since in the quark case we have:

\begin{eqnarray}
\label{quark}
{1\over{2M}}\int{{d\lambda \over2\pi}}e^{i\lambda x}
\langle p,s|\bar \psi (0)\gamma^{\mu}
\gamma ^5\psi (\lambda n)|p,s \rangle=g_1(x)(sn)p^{\mu}+g_T (x) s_T^{\mu}.
\end{eqnarray}

The quantity
$g_T$ was shown
to be the good variable to study the generalized Gerasimov-Drell-Hearn
sum rule, and the $x-$dependence of the
anomalous gluon contribution \cite{ST95}.
The latter result was recently confirmed \cite{rav}.

The light-cone distributions $\Delta G$ and $\Delta G_T$ can be
easily obtained
by the projection of gluon density matrix,so we have,

\begin{eqnarray}
\label{distr}
\Delta G(x)={1\over{4M(sn)}} \int{{d\lambda \over2\pi}}
e^{i\lambda x}
\langle p,s|A_\rho(0) A_\sigma (\lambda n) |p,s \rangle
\epsilon ^{\rho \sigma p n}, \nonumber \\
\Delta G_T(x)={1\over{4M(s^2)}} \int{{d\lambda \over2\pi}}
e^{i\lambda x}
\langle p,s|A_\rho(0) A_\sigma (\lambda n) |p,s \rangle
\epsilon ^{\rho \sigma p s}.
\end{eqnarray}

Now by making use of the axial gauge $An=0$, one may express their moments
\footnote {The first moment requires to take into account the non-local
operator \cite{ET88,BB}.At the same time, the
non-local operators found in the renormalization of
the gluon contribution to $g_2$ \cite{BET94}
should be equal to zero, when one uses the gauge invariance and
equations of motion.}
in terms of gluon field
strength $G_{\mu \nu}$, according to

\begin{eqnarray}
\label{distrG}
\int_0^1 dx x^k \Delta G(x)={1\over{4M(sn)}}
\langle p,s|G_{\rho n}(0) (i \partial n)^{k-1} G_{\sigma n}(0) |p,s \rangle
\epsilon ^{\rho \sigma p n}, \nonumber \\
\int_0^1 dx x^k \Delta G_T(x)={1\over{4M(s^2)}}
\langle p,s|G_{\rho n}(0) (i \partial n)^{k-1} G_{\sigma n}(0) |p,s \rangle
\epsilon ^{\rho \sigma p s}.
\end{eqnarray}

We denote here $G^{\mu n}=G^{\mu \nu} n_{\nu},
\partial n=n_{\mu}\partial^{\mu}$, and we recall that in configuration space
$x^k=(i \partial n)^k$.
The kinematical identities, implied by the vanishing of the
totally antisymmetric tensor of rank 5 in four-dimensional space,

\begin{eqnarray}
\label{kin}
n^{\mu}\epsilon_{\rho \sigma p n}-
n^{\sigma}\epsilon_{\rho \mu p n}+
n^{\rho}\epsilon_{\sigma \mu p n}=\epsilon_{\rho \sigma \mu n} \
\mbox{and}
\ n^{\mu}\epsilon_{\rho \sigma p s}-
n^{\sigma}\epsilon_{\rho \mu p s}+
n^{\rho}\epsilon_{\sigma \mu p s}=\epsilon_{\rho \sigma \mu s_T}
\end{eqnarray}
allow one to come to the standard gluonic operators,
used in the operator product expansion for spin-dependent
case \cite{AR}

\begin{eqnarray}
\label{momG}
\int_0^1 dx x^k \Delta G(x)=
{i^{k-1}\over{2M(sn)}}
\langle p,s|\tilde G_{\sigma \alpha}(0)
\partial ^{\mu_1}...\partial ^{\mu_{k-1}}
G_{\sigma \beta}(0) |p,s \rangle
n^{\alpha} n_{\beta} n_{\mu_1}...n_{\mu_{k-1}}, \nonumber \\
\int_0^1 dx x^k  \Delta G_T(x)=
{i^{k-1}\over{2M(s^2)}}
\langle p,s|\tilde G_{\sigma \alpha}(0)
\partial ^{\mu_1}...\partial ^{\mu_{k-1}}
G_{\sigma \beta}(0) |p,s \rangle
s_T^{ \alpha} n_{\beta} n_{\mu_1}...n_{\mu_{k-1}},
\end{eqnarray}
where $\tilde G_{\sigma \alpha}={1 \over 2} \epsilon_{\sigma \alpha \mu \nu}
G_{\mu \nu}$.
Taking the totally symmetric part of the matrix element

\begin{eqnarray}
\label{matr}
i^{k-1}\langle p,s|\tilde G_{\sigma \alpha}(0)
\partial ^{\mu_1}...\partial ^{\mu_{k-1}}
G_{\sigma \beta}(0) |p,s \rangle =a_k S_{\alpha \beta \mu_1...\mu_{k-1}}
s^{\alpha}
 p^{\beta}   p^{\mu_1}... p^{\mu_{k-1}},
\end{eqnarray}
where $S$ denotes the total symmetrization and $a_k$ is the scalar constant,
one immediately obtains the relation

\begin{eqnarray}
\label{momWW}
\int_0^1 dx x^k \Delta G(x)=(k+1)\int_0^1 dx x^k  \Delta G_T(x),
\end{eqnarray}
which is equivalent to the WW formula:

\begin{eqnarray}
\label{xWW}
\Delta G_T(x)=\int_x^1 {\Delta G (z)\over z}dz.
\end{eqnarray}

The existence of this relation is very natural because of the
similarity between quark  and gluon
density matrices (see eqs. (\ref{quark}) and (\ref{dm}) ).
Our present knowledge on $\Delta G(x)$, which is not very precise,
allows a great freedom, so several different parametrizations have been
proposed in the literature \cite{B1,B2,GS,GRSV}. We show in Figs 1a and 2a
some possible gluon helicity distributions $x \Delta G(x)$ and in Figs 1b and
2b the corresponding $x \Delta G_T(x)$ obtained by using (\ref{xWW}).
It is worth
noting from these pictures that, in all cases $x \Delta G(x)$ and
$x \Delta G_T(x)$ are rather similar in shape and magnitude.

Let us now move to the calculation of short-distance subprocess.
For this, it is instructive to compare the two terms in the
gluon density matrix (\ref{dm}).
While the longitudinal term is in fact a two-dimensional transverse
antisymmetric tensor and corresponds to the density matrix
of a circularly polarized gluon

\begin{eqnarray}
\label{londen}
\Delta G(x) \epsilon ^{\rho \sigma p n}=
\Delta G(x) \epsilon_{TT} ^{\rho \sigma} ,
\end{eqnarray}
the transverse polarization term generates the circular
polarization in the plane, defined by one transverse and one
longitudinal direction

\begin{eqnarray}
\label{trden}
M \Delta G_T(x) \epsilon ^{\rho \sigma s_T n}=
\Delta G_T(x) \epsilon_{TL} ^{\rho \sigma} ,
\end{eqnarray}
and therefore corresponds to the circular {\it transverse} polarization
of gluon. Such a polarization state is clearly impossible for
on-shell collinear gluons. They should have either nonzero virtuality,
or nonzero transverse momentum. Note that one of these effects is required to
have nonzero anomalous contribution to the first moment of the structure
function $g_1$\cite{EST,CCM}.
One may consider this similarity as supporting the
mentioned relations between $\Delta G_T$ and anomalous gluon contribution
\cite{ST95,rav}.

We should adopt the second possibility, namely a nonzero
transverse momentum, because the gluon remains on-shell
and the explicit gauge invariance is preserved. In this case,
the transverse polarization of nucleon may be converted to the longitudinal
circular polarization of gluon. The similar effect for quarks was
discussed earlier \cite{Ratcl,kt}.

To calculate now the asymmetry
in short-distance subprocess it is enough to find
the effective longitudinal polarization by projecting the
transverse polarization onto the gluon momentum:

\begin{eqnarray}
\label{pro}
s_L={\vec s_T \vec k \over{|\vec k|}} =s_T {k_T \over k_L}.
\end{eqnarray}

The partonic double transverse asymmetry can be easily obtained from the
longitudinal one according to,

\begin{eqnarray}
\label{att}
\hat A_{TT}={ {k_{T1} k_{T2}}\over {k_{L1} k_{L2}}} \hat A_{LL}.
\end{eqnarray}
By neglecting the transverse momentum dependence of $\hat A_{LL}$ one has

\begin{eqnarray}
\label{attq}
\hat A_{TT}={2\langle k_T^2\rangle \over {\hat s}} \hat A_{LL},
\end{eqnarray}
where $\hat s$ is the partonic c.m. energy.
Here we see that $\hat A_{TT}$ is strongly suppressed with respect to
$\hat A_{LL}$ and even more than in the case of partonic processes
with initial quarks and antiquarks \cite{JaSa}. If we consider
the hadronic double transverse asymmetry $A_{TT}$ for the two-jet production,
one can simply relate it, within some approximation, to the corresponding
double helicity asymmetry $A_{LL}$ as follows

\begin{eqnarray}
\label{atth}
A_{TT}={2\langle k_T^2\rangle \over M^2_{JJ}} {\Delta G_T (x_1) \over
\Delta G (x_1) }
{\Delta G_T (x_2)\over \Delta G (x_2) }A_{LL},
\end{eqnarray}
where $M_{JJ}$ is the invariant mass of the dijet.
Since $\Delta G_T (x)/ \Delta G (x)$ is of order of unity
and assuming the
average transverse momentum to be of the order of $1 GeV$,
we see that for  $M_{JJ}=10 GeV$, where $A_{LL}$ is at most $10\%$
at the maximum energy of RHIC, it
leads to $A_{TT} \sim 0.1\%$.
Actually, this small number is due to a large extent
to the small longitudinal asymmetry and small suppression factor
$k_T/M_{JJ}$. One may wonder that the double transverse asymmetry of
the direct photons with moderate $p_T \sim 5 GeV/c$ should be
of order $A_{TT} \sim A_{LL}  (k_T/p_T)^2 \sim 1\%$.

In conclusion, we have obtained the analog of the
Wandzura-Wilczek relation for
the transverse circular gluon polarization which has same order of
magnitude and sign as the gluon helicity distribution.
A transversely polarized nucleon
picks up the intrinsic transverse momentum of gluon
defining naturally the mass parameter of transverse asymmetries
for subprocesses with initial gluons (see eq.(\ref{attq})).
This leads to a strong suppression for the hadronic double transverse
asymmetry $A_{TT}$, which is expected to be $10^{-3}$ for
dijet production and $10^{-2}$ for direct photon production at RHIC energy.

We are indebted to V.M. Braun, A.V. Efremov and J.M. Virey 
for useful discussions.
O.T. is grateful to Centre de Physique Th\'eorique for warm hospitality
and to the Universit\'e de Provence for financial support. He was
partially supported by Russian Foundation of Fundamental Investigation
under Grant 96-02-17361.
This investigation was supported in part by INTAS Grant 93-1180.

\newpage

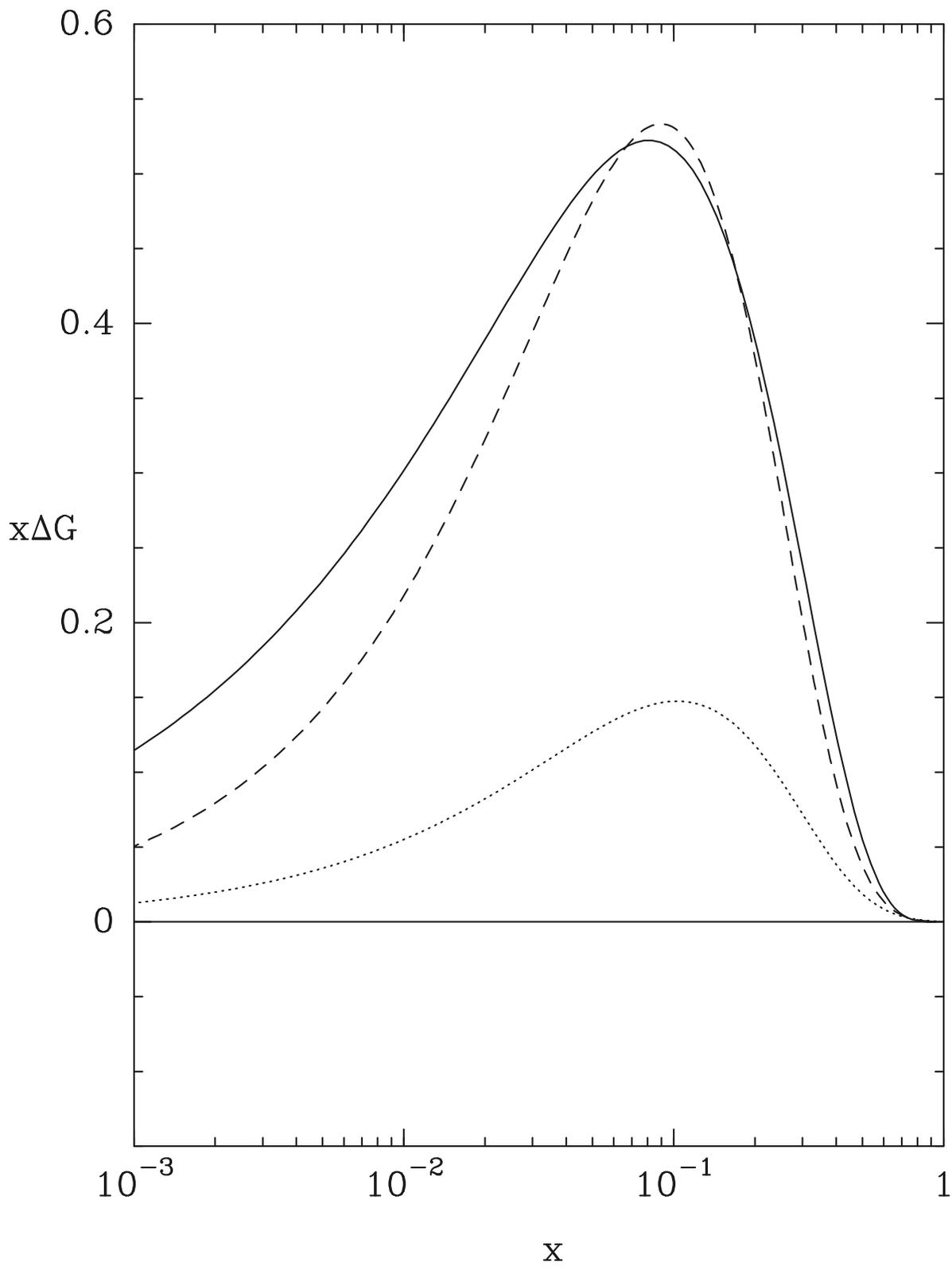
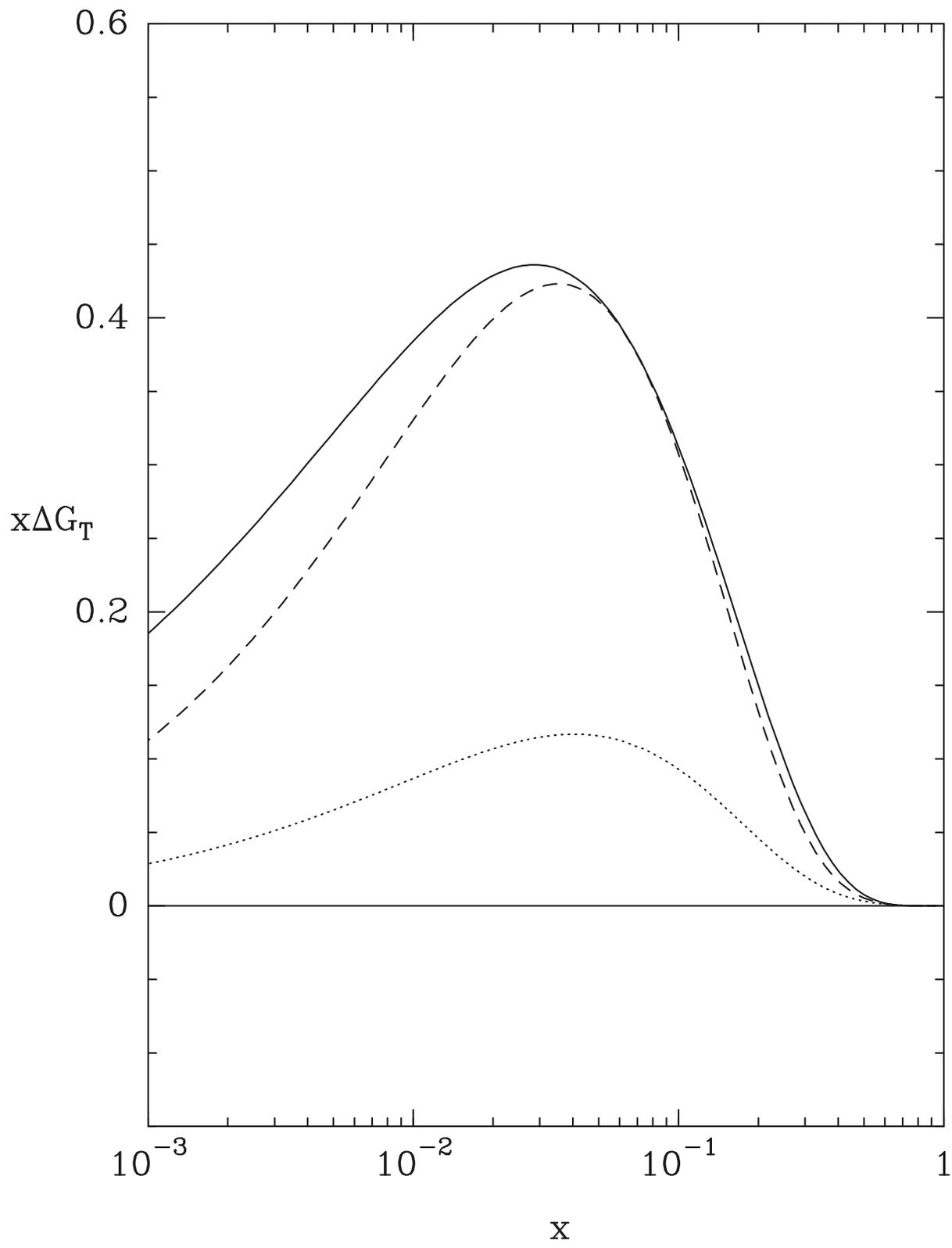
\begin{figure}[ht]
\hfill
\begin{minipage}{6.5in}
\caption{Fig. 1a. The gluon helicity distributions versus $x$ from
ref.\protect\cite{B1} (soft gluon is dotted curve, hard gluon
is dashed curve)
and from ref.
\protect\cite{B2} (solid curve).
Fig. 1b. The transverse polarization of the gluon obtained using
eq. (9) with labels corresponding to Fig. 1a}
\end{minipage}
\end{figure}

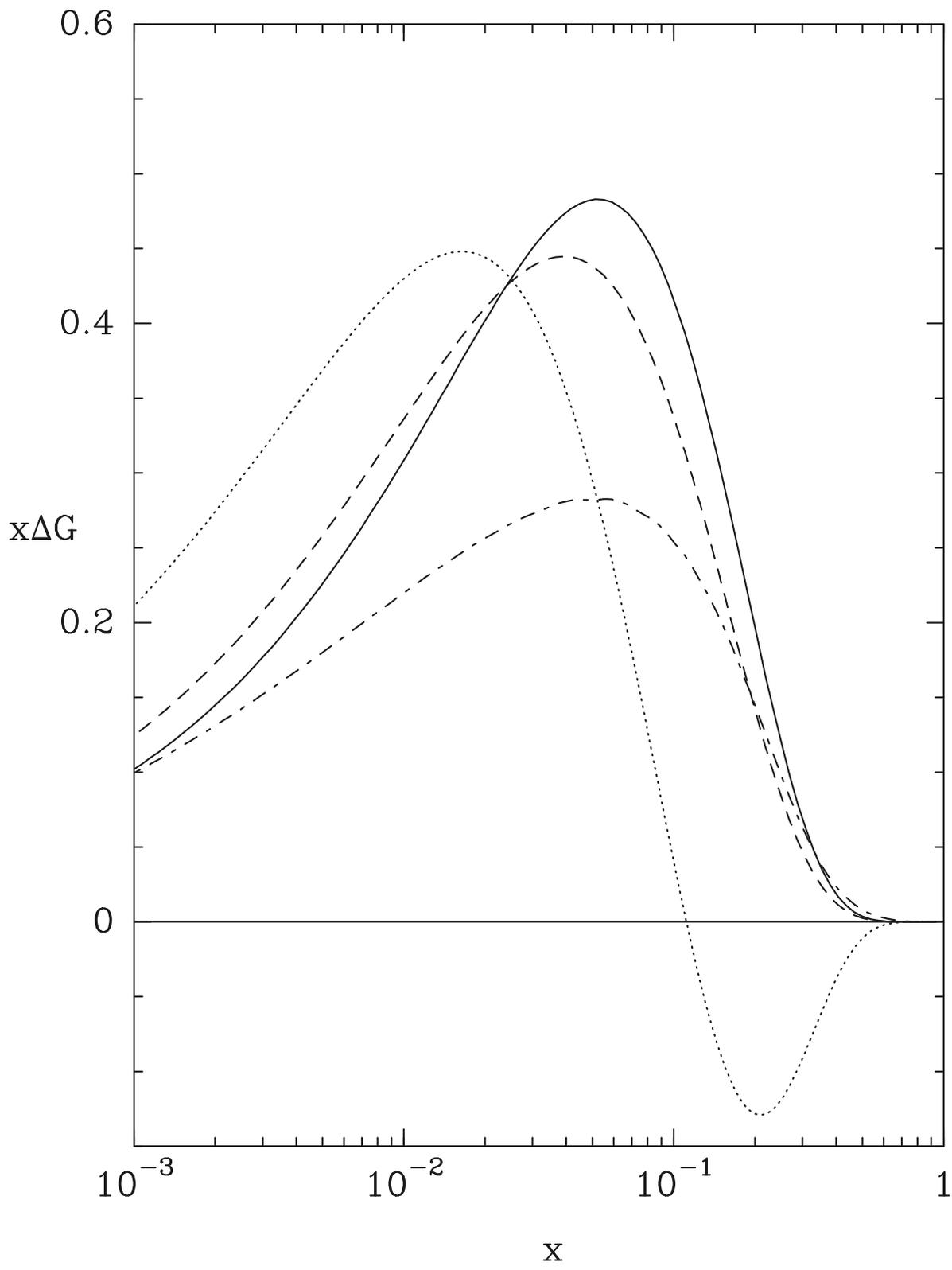
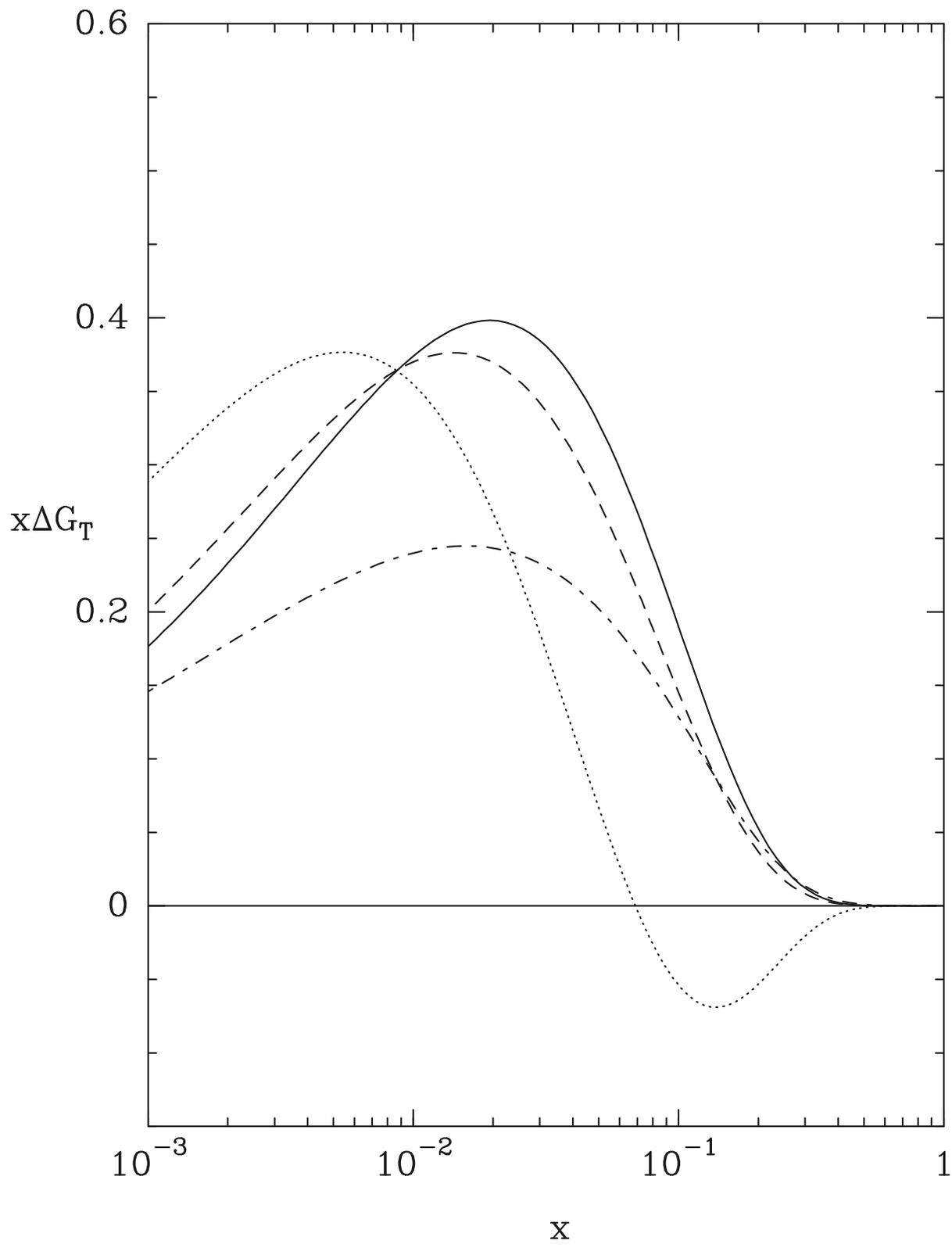
\begin{figure}[ht]
\hfill
\begin{minipage}{6.5in}
\label{fig2a}
\caption{Fig. 2a. The gluon helicity distributions versus $x$ from
ref.\protect\cite{GS} (gluon A is solid curve, gluon B
is dashed curve,  gluon C
is dotted curve )
and from ref.
\protect\cite{GRSV} (standard scenario, dotted dasched curve).
Fig. 2b. The transverse polarization of the gluon obtained using
eq. (9) with labels corresponding to Fig. 2a}
\end{minipage}
\end{figure}

\end{document}